\documentstyle[11pt]{article}
\normalbaselines
\begin{document}

\def\bbx{\begin{picture}(.4,.4) 
\thinlines
\put(.0,.4){\line(0,-1){.4}}
\put(.0,.4){\line(1,0){.4}}
\put(.4,0){\line(-1,0){.4}}
\put(.4,0){\line(0,1){.4}}
\end{picture}}

\unitlength 8mm
\def \mhrt{\begin{picture}(.3,.4)
\put(-.05,0.1){\mbox{${\,}_\heartsuit$}}
\end{picture}}

\def\ca#1{{\cal A}^{(#1)}}
\def\cA{{\cal A}}
\def\cB{{\cal B}}
\def\ab{{\bf a}}
\def\bb{{\bf b}}
\def\edd{\end{document}}
\def\tr{\mbox{{\rm Tr} }}
\newcommand{\sv}{\mbox{{\boldmath$ \sigma$ \unboldmath}}}
\newcommand{\cass}{\mbox{${\cal A}^{\hrt}$}} 
\newcommand{\cas}{\ifmmode{\cass}\else{\cass\ }\fi}
\newcommand{\capp}{\mbox{${\cal A}^{\hrt}_+$}}
\newcommand{\casp}{\ifmmode{\capp}\else{\capp\ }\fi}
\newcommand{\can}{\mbox{${\cal A}^{\hrt}_-$}}
\newcommand{\casn}{\ifmmode{\can}\else{\can\ }\fi}%
\def\brcc{\left(\begin{array}{cc}} 
\def\brrr{\left(\begin{array}{rr}} 
\def\brll{\left(\begin{array}{ll}} 
\def\eerr{\end{array}\right)}
\def\ttc#1#2#3#4{\brcc{#1}&{#2}\\{#3}&{#4}\eerr}
\def\ttm#1#2#3#4{\brcc{#1}&{#2}\\{#3}&{#4}\eerr}
\def\ttr#1#2#3#4{\brrr{#1}&{#2}\\{#3}&{#4}\eerr}
\def\ttl#1#2#3#4{\brll{#1}&{#2}\\{#3}&{#4}\eerr}

\def\ra{\rightarrow}
\def\b#1#2{\left(\nsp \begin{array}{c}#1\\#2\end{array} \nsp \right)}
\def\hr#1#2{\left[\nsp \begin{array}{c} #1\\ #2 \end{array} \nsp
\right]}
\newcommand{\alb}{{\mbox{\boldmath $\alpha$}}}
\newcommand{\anb}{{\mbox{$A_N$}}}
\newcommand{\an}{\ifmmode{\anb}\else{\anb\ }\fi} 
\newcommand{\snb}{{\mbox{$S_N$}}}
\newcommand{\sn}{\ifmmode{\snb}\else{\snb\ }\fi} 
\newcommand{\gfb}{{\mbox{$\Gamma^f$}}}
\newcommand{\gf}{\ifmmode{\gfb}\else{\gfb\ }\fi} 
\newcommand{\gpb}{{\mbox{$\Gamma^P$}}}
\newcommand{\diag}{\mbox{diag}}
\newcommand{\gp}{\ifmmode{\gpb}\else{\gpb\ }\fi} 
\newcommand{\hf}{\frac 1 2}
\newcommand{\fr}{fundamental representation}
\newcommand{\ir}{irreducible representation}
\newcommand{\irs}{irreducible representations}
\newcommand{\sump}{\sum_{\pi \in \sn} }
\newcommand{\sqn}{\sqrt{N}}
\newcommand{\all}{\quad \forall}
\newcommand{\al}{\alpha}
\newcommand{\RR}{{I\!\!R}}
\newcommand{\CC}{{\>l\!\!\!C}} 
\newcommand{\A}{{\cal A}} 
\newcommand{\Pol}{{\cal A}} 
\newcommand{\bder}{\bar\partial}
\newcommand{\id}{{\rm id}}
\newcommand{\Def}[1]{{\it #1}}
\newcommand{\scalar}[2]{#1\!\cdot\! #2}
\newcommand{\cross}[3]{[#1\!{\times}\! #2]_#3}


\newcommand{\nc}{\newcommand}
\nc{\bi}[2]{\left(\! \! \begin{array}{c}#1 \\ #2\end{array} \! \!
\right)}
\def\R{I\!\! R}
\def\t{\tau}
\def\ph{\varphi}
\def\th{\theta}
\newcommand{\la}{\lambda}
\newcommand{\xu}{\chi^{(U)}}
\def\gb{{\bf g}}
\def\sp #1{\langle #1 \rangle}
\newcommand{\s}{\sigma}
\newcommand{\prop}[1]{\noindent {\bf Proposition #1~:~}}
\newcommand{\pf}{\noindent {\em Proof}~:~~}
\newcommand{\be}{\begin{equation}}
\newcommand{\ee}{\end{equation}}
\newcommand{\el}[1]{\label{#1} \end{equation}}
\newcommand{\erl}[1]{\label{#1} \end{eqnarray}}
\newcommand{\lb}[1]{\label{#1}}
\newcommand{\rf}[1]{(\ref{#1})}
\newcommand{\br}{\begin{eqnarray}}
\newcommand{\eq}{\!\!\!\! &=& \! \! \!\!}
\newcommand{\eqq}{\!\!\!\!\! &=& \! \! \!\!}
\newcommand{\er}{\end{eqnarray}}
\newcommand{\nn}{\nonumber}
\newcommand{\for}{\qquad \mbox{ for}\quad}
\newcommand{\aand}{\qquad \mbox{ and}\quad}
\newcommand{\where}{\quad \mbox{ where}\quad}
\newcommand{\with}{\quad \mbox{ with}\quad}
\newcommand{\orr}{\qquad \mbox{ or}\quad}
\newcommand{\hrt}{\heartsuit}
\newcommand{\qed}{\hfill{\bbx}}

\def\x{\times}
\def\zb{{\bf b}}
\def\gbd{{\bf g}^D}
\def\gbdz{{\bf g}^D_0}
\def\ca#1{{\cal A}^{(#1)}}
\def\ad{\mbox{ad~}}
\def\cj{{\cal J}}
\def\di{\diamond}

\def\vline{\begin{picture}(.4,.5) 
\thinlines
\put(.1,-.5){\line(0,1){1.2}}
\end{picture}}

\def\eh{\hat{e}}
\def\wbar#1{\overline{#1}} 
\def\Om{\Omega}
\def\ga{\gamma}
\def\Omh{\widehat{\Omega}}
\def\laz{\lambda_0}
\def\cq{{\cal Q}} 
\def\co{{\cal O}} 
\def\Lra{\Longleftrightarrow}
\def\lra{\leftrightarrow}
\def\Ra{\longrightarrow}
\def\wb#1{\overline{#1}} 



\begin{center}
{\LARGE \bf
Matrix Representation of Octonions and
Generalizations} \\ [5mm]

Jamil \ Daboul
{\footnote
{\it Permanent address: Physics Department, Ben
Gurion University of the Negev,
84105 Beer Sheva, ISRAEL\\
(E-mail: daboul@bgumail.bgu.ac.il)}}
and Robert Delbourgo
{\footnote
{\it E-mail: Bob.Delbourgo@utas.edu.au}}
\\[5mm]
{\it School of Mathematics and Physics,
University of Tasmania \\
Hobart, Australia}\\[3mm]
\end{center}

\begin{abstract}
We define a special matrix multiplication among
a special subset of $2N\x 2N$ matrices, and study the
resulting (non-associative) algebras and their
subalgebras. We derive the conditions under which
these algebras become alternative non-associative
and when they become associative. In
particular, these algebras yield special matrix
representations of octonions and complex numbers;
they naturally lead to the Cayley-Dickson doubling process.
Our matrix representation of octonions also yields
elegant insights into Dirac's equation for a free particle.
A few other results and remarks arise as byproducts.
\end{abstract}

\section*{I. Introduction}

Complex numbers and functions have played a pivotal role in physics for
three centuries. On the other hand, their generalization to other
Hurwitz
algebras does not seem to have fired the interest of physicists to the
same
extent, because there is still no {\em compelling} application of them.
Thus, despite the fascination of quaternions and octonions for over a
century, it is fair to say that they still await universal acceptance.
This
is not to say that there have not been valiant attempts to find
appropriate
uses for them. One can point to their possible impact on
\begin{itemize}
\item Quantum mechanics and Hilbert space \cite{QM}
\item Relativity and the conformal group \cite{Rel}
\item Field theory and functional integrals \cite{QFT}
\item Internal symmetries in particle physics \cite{GUT}
\item Colour field theories \cite{Colour}
\item Formulations of wave equations \cite{WE}
\end{itemize}
In all these cases, there is nothing that stands out and commands our
attention; rather, the attempts to describe relativistic physics in
terms of quaternions and octonions look rather contrived if not
forced, especially for the case of octonions. In this paper we
describe a generalization of octonions that allows for Lie algebras
beyond the obvious SU(2) structure that is connected with quaternions.
We do not presume that they will lead to new physics, but we do think
they will at least provide a new avenue for investigation.

Since octonions are not associative, they cannot be
represented by matrices with the usual
multiplication rules. In this note,
we give representations of octonions and
other non-associative algebras by special matrices,
which are endowed with very special
multiplication rules; these rules can be regarded
as an adaptation and generalization of Zorn's
multiplication rule \cite{zorn}. These matrix
representations suggest generalizations of octonions to other
non-associative algebras, which in turn lead one
almost automatically to a construction of
new algebras from old ones, with double the
number of elements; we have called these `double algebras'.
Closer inspection reveals that our procedure can be
made to correspond to the Cayley-Dickson
construction method \cite{schaf}, except that in our
case the procedure seems rather natural, once one accepts the
multiplication rule, whereas the Cayley-Dickson rule
looks {\em ad hoc} at first sight.

\section*{II. Definitions, notations and a review of the
octonion algebra $\co$}

The Cayley or the octonion algebra $\co$ is an
8-dimensional non-associative algebra, which is
defined in terms of the basis elements
$e_\mu~~(\mu=0,1,2,\cdots,7)$ and their
multiplication table. $e_0$ stands for the
unit element. We can efficiently summarize the table by
introducing  the following notation [in general, we shall use
Greek indices $(\mu,\nu,\ldots)$ to include the $0$ and
latin indices $(i,j,k,\ldots)$ when we exclude the
$0$]:
\be
\hat e_k \equiv e_{4+k}~, \for k = 1,2,3~.
\ee
The multiplication rules among the basis elements of
octonions $e_\mu$ can be expressed in the form:
\br
-e_4 e_i =e_ie_4 &=&\eh_i ~, \qquad
e_4 \eh_i =-\eh_i e_4 =e_i~,
\qquad e_4 e_4 =-e_0~, \\
e_i e_j &=&-\delta_{ij}~e_0 ~+~\epsilon_{ijk}
e_k~, \lb{quat} \\
\eh_i \eh_j &=&- \delta_{ij}~e_0 ~-~\epsilon_{ijk}
e_k~, \qquad \qquad (i,j,k=1,2,3)
\\
-\eh_j e_i= e_i \eh_j &=& -\delta_{ij}~e_4 ~-~
\epsilon_{ijk}\eh_k~.
\er
We can formally summarize the rules above by
\be
e_\mu ~e_\nu = g_{\mu\nu}~ e_0 ~+~
\sum_{k=1}^7 \ga_{\mu\nu}^k e_k~,
\quad g_{\mu\nu}:={\rm diag~}(1,-1,\cdots,-1)~,
\quad \ga_{ij}^k= -\ga_{ji}^k~,~~
\el{gmr}
where $\mu,\nu=0,1,\cdots,7~,$ and
$i,j,k=1,\cdots,7~$.
The multiplication properties are sometimes displayed
graphically by a circle surrounded by a
triangle, but we shall not bother to exhibit that.

The multiplication law \rf{quat} shows that the
first four elements form a closed {\em
associative} subalgebra of $\co$, which is known as
the {\em quaternion algebra},
\be
\cq \equiv
\sp{e_0,e_1,e_2,e_3}_\RR~.
\ee
while the other rules (2), (4) and (5) show that $\co$ can
be graded as follows:
\be
\co = \cq \oplus \hat \cq~,
\where \hat \cq :=e_4 \cq~.
\el{grad}
$\co$ is a non-associative algebra. Now a measure
of the non-associativity in any algebra $\cA$ is
provided by the {\em associator}, which is defined
for any 3 elements, as follows
\be
(x,y,z) := (xy)z- x(yz)~, \for x,y,z \in \cA~.
\ee
In particular, the associators for the octonion basis are
\be
(e_i,e_j,e_k)=2 \epsilon_{ijkl} e_l~,
\ee
where $\epsilon_{ijkl}$ are {\em totally}
antisymmetric
\cite{f1}
and equal to unity for the following 7
combinations \cite{Colour}:
\be
1247,~1265,~2345~,2376~,3146~,3157~\mbox{and}
~4576~.
\ee

\subsection*{The quaternionic subalgebra $\cq$}

It is very well-known that the quaternions form
an associative subalgebra $\cq$~,
which can be represented by the Pauli matrices:
\be
e_0\ra \s_0=1~,~~
\aand e_j \ra -i\s_j~~~(j=1,2,3)~,
\ee
where, as usual,
\be
\s_0=\ttl{1}{0}{0}{1}~,~~
\s_1=\ttl{0}{1}{1}{0}~,~~
\s_2=\ttl{0}{-i}{i}{0}~,~~
\s_3=\ttl{1}{0}{0}{-1}~.
\ee
It is trivial to check that the above map is an isomorphism:
\be
e_ie_j \Lra -\s_i\s_j=
-(\delta_{ij}+i\epsilon_{ijk}~\s_k)\Lra
-\delta_{ij}+\epsilon_{ijk} ~e_k~.
\ee

\section*{III. Non-associative multiplication}

In contrast to $\cq$, {\em the Cayley algebra
$\co$ cannot be represented by matrices with the
usual multiplication rules,} because $\co$ is not
associative. However, as we demonstrate below, it is possible
to represent octonions by matrices, provided one defines a
special multiplication rule among them.

\subsection*{Zorn's representation of octonions}

Zorn \cite{zorn} gave a representation of the
octonions \cite{schaf} in terms
of $2 \x 2$ matrices $M$, whose diagonal
elements are scalars and whose off-diagonal
elements are 3-dimensional vectors:
\be
\co \ni x \Ra \ttc{\al}{\ab}{\bb}{\beta}~,
\ee
and invoked a peculiar multiplication rule for these
matrices \cite{zorn}. With slight modification of the rule
adopted by Humphreys \cite{hum} p. 105 our rule is:
\be
{\ttc{\al} {\ab} {\bb} {\beta}} *
{\ttc{\al' }{\ab' }{\bb' }{\beta' }}=
\ttl{\al\al'+ \ab \cdot \bb'}
{~\al \ab'+ \beta' \ab -\bb\x \bb'}
{\al' \bb+ \beta \bb'+\ab \x \ab'}
{~\beta\beta'+\bb\cdot\ab'}~.
\el{zmr}
We propose to adapt this multiplication law to octonions and
also replace
the necessary 3-dimensional basis
vectors $\hat v_k$ by Pauli
matrices $\s_k~(k=1,2,3)$, so that the octonions
can be represented by the following ordinary $4\x 4$
matrices:
\be
\begin{array}{ccc}
e_0~\Longleftrightarrow \Om_0\equiv
\ttm{1}{0}{0}{1}~, &
\quad e_k\Longleftrightarrow \Om_k\equiv
\ttm{0}{-\s_k}{\s_k}{0}~,
&\quad (k=1,2,3) \cr
~&~& \cr
e_4~\Longleftrightarrow \Om_4\equiv
\ttm{i}{0}{0}{-i}~, & \qquad
\hat e_k\Longleftrightarrow \Omh_k \equiv
\ttm{0}{i\s_k}{i\s_k}{0}~. &~
\end{array}
\el{octr}
(Note the equality of $\Om_k ~~(k=1,2,3)$ to the Dirac
matrices $\gamma_k$~, and $\Om_4$ to $i\gamma_0$ in the
Pauli-Dirac representation.)
It can be shown by explicit multiplication, that
the above map \rf{octr} becomes an isomorphism,
provided we define the modified product rule, which we denote by
$\mhrt$ :
\be
\ttc{\al} {A} {B} {\beta} \mhrt
\ttc{\al' }{A' }{B' }{\beta' }=
\ttl{\al\al'+ \frac 1 2\tr (AB')}
{~~\al A'+ \beta' A+\frac i 2 [B,B']}
{\al' B+ \beta B'-\frac i 2 [A,A']}
{~~\beta\beta'+\frac 1 2\tr(BA')}~.
\el{mroct}
where $[A,B]\equiv AB-BA$ ~is the commutator of $A$
and $B$~. Of course,
$A=\ab\cdot \sv$ and $B=\bb \cdot \sv$ are
traceless: ~ $\tr A=\tr B=0$~.\\

In particular, the above
multiplication rule yields the following relations
\br
\ttm{0}{\eta \s_i}{\xi\s_i}{0}
~\mhrt~ \ttm{0}{\eta' \s_j}{\xi' \s_j}{0}
&=&\ttl{\eta\xi'\delta_{ij}}
{~\xi\xi' \frac i 2 [\s_i,\s_j]}
{-\eta\eta' \frac i 2 [\s_i,\s_j]}
{~\xi \eta'\delta_{ij}} \nn \\
&=&
\delta_{ij} \ttm{\eta\xi'}{0}{0}{\xi \eta'}
+\epsilon_{ijk}~ \ttm{0}{-\xi\xi'\s_k}
{\eta\eta'\s_k}{0}~,
\er
which are helpful for checking the multiplication
rules (2)-(5), by substituting the appropriate
coefficients, $\eta$ and $\xi$~.

\subsection*{The standard conjugate of octonions}

Usually, octonions are studied over the field of
real numbers $\RR$,
\be
 x = \sum_{\mu=0}^7 x_\mu e_\mu \equiv x_0 + {\bf x},
 \qquad \for x_\mu \in \RR,
\ee
although later we will find it
interesting to deal with their complex extension.
The standard {\em conjugate} $\bar x$ of an octonion over
$\RR$ is defined by
\be
\bar x := x_0 e_0 -\sum_{i=1}^7 x_i e_i
\equiv x_0 -{\bf x}.
\ee
The reason for this definition is that the product of $\bar x$
with $x$ yields a positive definite norm:
\be
n(x)= x\bar x=\bar x x=\sum_{\mu=0}^7 x_\mu^2 \ge 0~.
\ee
Moreover, this norm obeys the composition law,
\be
n(xy) = n(x)n(y).
\ee
However with complex octonions (real $x \rightarrow$ complex $z$ in
(20)) we shall still formally define
the {\em conjugate} $\bar z$ of $z$, to be
\be
\bar z := z_0 e_0 -\sum_{i=1}^7 z_i e_i~,
\for z_\mu \in \CC~.
\ee
It follows that the product $z \bar z$ is again
proportional to unity~:
\be
n(z)= z \bar z= \bar z z=
\sum_{\mu=0}^7 z_\mu^2 \in \CC~,
\el{znorm}
but $n(z)$ ceases to be real in general; therefore $n(z)$
should simply be regarded as a scalar function, but not a norm.

It is interesting to calculate $n(z)$ by using
the matrix representation \rf{matr}:
Firstly, we note that if $z$ is mapped into
the matrix $Z$, then $\bar
z$ will be mapped into $\bar Z$, as follows:
\be
z\longrightarrow
Z\equiv \sum_{\mu=0}^7 z_\mu\Om_\mu=
\ttc{\al}{A}{B}{\beta}~,
\qquad \bar z \longrightarrow
\bar Z \equiv \ttc{\beta}{-A}{-B}{\al}~,
\el{matr}
where $A=\ab\cdot \sv$ and $B=\bb\cdot \sv$,
with
\be
\alpha= z_0+iz_4~,~~\beta=z_0-iz_4~,~~
a_k= -z_k+iz_{4+k}~,~~
b_k= z_k+iz_{4+k}~~ (k=1,2,3).
\el{coef}
Secondly,
\be
z\bar{z}\lra
Z \mhrt \wbar{Z} =\ttc{\al}{A}{B}{\beta}\mhrt
\ttc{\beta}{-A}{-B}{\al}
= (\al\beta - \hf \tr AB)
\ttc{1}{0}{0}{1} =n(z)~ I_{4\x4}
\el{ml}
Therefore, we reproduce the expression \rf{znorm},
as expected:
\be
n(z) := \frac 1 4 \tr (Z \mhrt \wbar {Z})
= \al\beta - \hf \tr AB
= \al\beta - \ab\cdot \bb=
\sum_{\mu=0}^7 z_\mu^2 ~.
\ee

\subsection*{Hermitian conjugate of octonions}

Since $\s_i$ are Hermitian matrices,
{\em all our representation matrices
$\Om_k$ are anti-hermitian, with the exception of
the identity $\Om_0$} (which is Hermitian of course)~:
\be
\Om_k^\dag = - \Om_k~,\qquad k=1,2,\cdots, 7~.
\ee
This fact enables us to prove that the
following `hermiticity' property also
holds for the $\mhrt$ products:
\be
(\Om_\mu \mhrt\, \Om_\nu)^\dag =
\Om_\nu^\dag \mhrt\, \Om_\mu^\dag~, \for
\mu,\nu=0,1,\dots,7~.
\el{np}
First, we note that this equality holds trivially
for
$(\Om_0 \mhrt\, \Om_\mu)^\dag = \Om_\mu^\dag =
\Om_\mu^\dag \mhrt\, \Om_0^\dag $.
Second, we prove \rf{np} for $j, k\ne 0$ by
using \rf{gmr} and noting that $\ga_{ij}^k$ are real and
antisymmetric in $j,k$, so that
\be
(\Om_j \mhrt\, \Om_k)^\dag =
-\delta_{jk}~ \Om_0 + \sum_{i=1}^7 \ga_{jk}^i \Om_i^\dag=
-\delta_{kj}~ \Om_0 +\sum_{i=1}^7 \ga_{kj}^i \Om_i=
\Om_k \mhrt\, \Om_j=
\Om_k^\dag \mhrt\, \Om_j^\dag ~,~~ j,k=1,\cdots,7~.
\ee
The conjugation property \rf{np} of
$\Om_\mu$ suggests the following formal
definition for the {\em Hermitian conjugate}
of the octonionic basis:
\be
e_0^\dag=e_0~, \qquad
e_j^\dag=-e_j~ ~~(j=1,2,\cdots,7)~,
\ee
whereupon the {\em `number operators'} become equal to
the identity element:
\be
N_\mu := e_\mu^\dag e_\mu =
e_\mu e_\mu^\dag =
e_0=1~, \quad
\mbox{\rm (no summation)}\quad ~~(\mu=0,1,\cdots, 7)~.
\ee
We can now define the {\em Hermitian conjugate} of
the complex octonions $z$ in a natural way, by
\be
z^\dag := \sum_{\mu=0}^7 \bar z_\mu e_\mu^\dag=
\bar z_0 e_0 - \sum_{i=1}^7 \bar z_i e_i \equiv
\bar z_0 - {\bar{\bf z}}~, {\rm where~} z_i \in \CC~.
\ee
We then calculate
\br
z z^\dag &\equiv&
(z_0 + {\bf z})(\bar z_0 - \bar{\bf z})=|z_0|^2 +
(\bar z_0 {\bf z} -z_0 \bar{\bf z}) - {\bf z}\bar{\bf z} \nn \\
&=& \sum_{\mu=0}^7 |z_\mu|^2
- \sum_{k=1}^7 (z_0 \bar z_k-z_k \bar z_0) e_k
+ \sum_{1\leq i < j}^7 (z_i \bar z_j-z_j \bar z_i) e_ie_j \nn \\
&=&
N(z) + 2i
\sum_{k=1}^7 \Im \left (\bar z_0 z_k +
\sum_{1\le i<j \le 7} z_i \bar z_j
\ga_{ij}^k \right)
e_k~,
\erl{zz}
where
\be
N(z) = \sum_{\mu=0}^7 |z_\mu|^2~.
\el{normz}
The definition $N(z)$ is perfectly reasonable for a norm
although the decomposition law (23) is {\em not} satisfied.
We see that the `space components' $(zz^\dag)_i$
of $zz^\dag$ are pure imaginary. To understand why this is
expected on general grounds, it is useful to
introduce the concept of a {\em Hermitian
octonion}: $y^\dag=y$~, which signifies that
\be
\bar y_0=y_0~, \quad \bar y_i=-y_i
~~(i=1,\cdots,7)~,
\ee
so that $y_0$ must be real and all the
`space components' must be pure imaginary.

Since $zz^\dag$ is Hermitian by (31), we see
that its space components can only be pure imaginary.
If we wish to get rid of these components and retain only the
zero component, we must add the standard conjugate.
Thus, we may define the {\em Hermitian norm} by
\be
N(z)=(zz^\dag+\wbar{zz^\dag})/2~.
\ee
Hence, if $z$ is mapped into $Z$, then
$z^\dag$ will be mapped into $Z^\dag$,
which is obtained by the standard Hermitian
conjugation of the matrix $Z$.

One of the main insights gained by using the matrix representation is
when we calculate the Hermitian norm.  If
\be
z\longrightarrow
Z= z^\mu\Om_\mu= \ttc{\al}{A}{B}{\beta}~,
\qquad {\rm then~~} z^\dag \longrightarrow
Z^\dag
=\ttc{\bar \al}{B^\dag}{A^\dag}{\bar \beta}~.
\el{matrh}
The product
\br
Z\mhrt Z^\dag &=&
\ttc{\al}{A}{B}{\beta} \mhrt
\ttc{\bar \al}{B^\dag}{A^\dag}{\bar \beta} \nn \\
&=&
\ttc{\al \bar \al+ \hf \tr AA^\dag}
{\al B^\dag+\bar \beta A+\frac i 2 [B,A^\dag]}
{\bar \al B + \beta A^\dag-\frac i 2 [A,B^\dag]}
{\beta \bar \beta+ \hf \tr BB^\dag}~.
\er
The zero component of $zz^\dag$ is proportional to the trace of
$Z\mhrt Z^\dag$, so that the new {\em Hermitian norm} can be
expressed in terms of the representation matrices $Z$ as follows:
\br
N(z) &=& \frac{1}{4} \tr (Z\mhrt Z^\dag) =
\hf \left(|\al|^2 + |\beta|^2\right) + \frac{1}{4} \left(\tr (AA^\dag)
+ \tr (BB^\dag) \right) \nn \\
&=&
\hf \left(|\al|^2+ |\beta|^2+
\sum_{k=1}^3 (|a_k|^2+|b_k|^2)\right) =
\sum_{\mu=0}^7 |z_\mu|^2~,
\er
in accordance with \rf{normz}.
For real $z_\mu$~ we get $\beta\rightarrow\bar \al~,
B\rightarrow -A^\dag~$ in (26). Therefore,
$ AB \rightarrow -AA^\dag= -{\bf a}\cdot \sv~ \bar{\bf a} \cdot
\sv = - {\bf a} \cdot \bar{\bf a}.$
Thus, the formally defined scalar reduces to a
conventional norm:
\be
N(z) = \al\beta-A\cdot B=|\al|^2+ {\bf a}\cdot \bar{\bf a}
\rightarrow |\al |^2 +\sum_{k=1}^3 |a_k|^2=
\sum_{\mu=0}^7 x_\mu^2 \equiv n(z)\ge 0 ~.
\ee

\subsection*{Non-associative algebras from Lie
algebras}

The main advantage of our matrix representation
over the Zorn vector representation, is that
our multiplication rule can be generalized to
{\em any number $n$ of dimensions},
whereas the Zorn rule is restricted,
since it is defined in terms of vector product
$\ab\x \bb $, which only applies to 3-vectors!

In particular, given {\em any} representation of
an $n$-dimensional Lie algebra $\gb~$ in terms of
Hermitian $N\x N$ matrices $\la_k~ (k=1,2,\cdots,n)~$,
we can then define $2n+2$ different $2N$-dimensional
matrices,
\be
\begin{array}{ccc}
e_0~\Longleftrightarrow \Om_0\equiv
\ttm{1}{0}{0}{1}~, &
\quad e_k\Longleftrightarrow \Om_k\equiv
\ttm{0}{-\la_k}{\la_k}{0}~,
&(k=1,\cdots,n) \cr
~&~& \cr
\hat e_0~\Longleftrightarrow \Om_{n+2}\equiv
\ttm{i}{0}{0}{-i}~, & \qquad
\hat e_k\Longleftrightarrow \hat \Om_k \equiv
\ttm{0}{i\la_k}{i\la_k}{0}~. &
\end{array}
\el{octrn}
If we multiply these matrices using the $\mhrt$
rule, we end up with a closed
algebra, which we shall call the {\em double
algebra $\gb^D$}, with the following product
rules for their basis elements
($\Omh_0 \equiv \Om_{n+2}$):
\br
-\Omh_0 \Om_k &=& \Om_k \Omh_0=\Omh_k~, \quad
\Omh_0 \Omh_k =-\Omh_k \Omh_0 =\Om_k~,
\quad \Omh_0\Omh_0 =-\Om_0~,
\nn \\
\Om_i \Om_j &=&- \delta_{ij}~ \Om_0 + f_{ijk}~
\Om_k~, \nn \\
\Omh_i \Omh_j &=&- \delta_{ij}~\Om_0-f_{ijk}~
\Om_k~, \lb{ggn} \\
-\Omh_j \Om_i= \Om_i \Omh_j
&=& -\delta_{ij}~ \Omh_0- f_{ijk}~\Omh_k~. \nn
\er
Above, the $f_{ijk}$ are the structure constants of
the Lie algebra $\gb$, defined as usual by
\be
[L_i,L_j]=if_{ijk} L_k~.
\ee
The matrices (44) can be regarded as the $\mhrt$-
matrix representation of the following
(non-associative) abstract algebra:
\br
-\eh_0 e_k &=&e_k \eh_0=\eh_k~, \quad
\eh_0 \eh_k =-\eh_k \eh_0 =e_k~,
\quad \eh_0\eh_0 =-e_0~, \where
\eh_0 \equiv e_{n+2}~, \lb{g1} \\
e_i e_j &=&- \delta_{ij}~ e_0 +f_{ijk} e_k~,
\\
\eh_i \eh_j &=&- \delta_{ij}~e_0-f_{ijk} e_k~,
\\
-\eh_j e_i= e_i \eh_j &=& -\delta_{ij}~ \eh_0-
f_{ijk}\eh_k~.
\erl{g2}
These rules \rf{g1}-\rf{g2} can all be summarized by
($\mu,\nu=0,1,2,\cdots,2n+2$)
\be
e_\mu e_\nu = g_{\mu\nu}~ e_0 ~+~
\sum_{k=1}^{2n+2} \ga_{\mu\nu}^k~ e_k,
\quad g_{\mu\nu}:={\rm diag~}(1,-1,\cdots,-1)~,
\quad \ga_{ij}^k= -\ga_{ji}^k~.
\el{mrg}

We note from (44) that the
$e_\mu~,~\mu=0,1,\cdots,n$ correspond to a
{\em subalgebra} $\gb_+$ of $\gb^D$.
The rules (47) show that the double algebra
$\gb^D$ is obtained from $\gb_+$ simply
by adding a new element, called $\hat
e_0$, and defining the other $\hat e_k$.
This Lie algebra example then
automatically leads us to a more general doubling procedure,
which can be applied to {\em any algebra} and not just to
those constructed from Lie algebras. In fact this doubling
idea is exactly the procedure which is known as the {\em
Cayley-Dickson process}, as we shall see below.

\section*{IV. Deformed multiplication and the ${\cal A}^\hrt$ algebra}

Begin with the following subset of
$2N\x 2N$-matrices:
\be
\cA := \left\{ {\ttc{\al}{A}{B}{\beta}}
~\vline
A, B \in M_{N\x N} \right\} ~,
\ee
where the $N\x N$ matrices $\al$ and $\beta$ in
the 1st and 4th
quadrants are proportional to unit matrices.

Among these matrices we may define a more general \cite{zsss}
multiplication rule than that given in \rf{mroct}.
We shall still denote it by $\mhrt$ since it only
introduces two complex {\em deformation parameters} $\la_0$
and $\la$ (their values will be restricted as we impose
further conditions on the subalgebras) :
\be
X ~\mhrt~ X'\equiv {\ttc{\al} {A} {B} {\beta}} \mhrt
{\ttc{\al' }{A' }{B' }{\beta' }} :=
{\ttl{\al\al'+ \laz~ A\cdot B'}
{~~\al A'+ \beta' A-\la [B,B']}
{\al' B+ \beta B'+\la [A,A']}
{~~\beta\beta'+ \laz~ B\cdot A'}}.
\el{mr}
As before, $[A,B]\equiv AB-BA~$ denotes the
commutator, but $A\cdot B $ may now be chosen to be any
suitable {\em bilinear map} into
an appropriate field $F$. For example, one might define
$A\cdot B $ by $A\cdot B \equiv \tr(AB)/N~$, or
if $A$ and $B$ belong to a Lie algebra,
then one could take $A\cdot B $ to be the adjoint trace:
$ A\cdot B:= \tr (\ad A \,\ad B)$~, where $\ad$
denotes the adjoint representation \cite{f2}

When $\la=0$ and $\la_0=1$ the
multiplication rule (53) looks {\em almost}
like the usual one for matrices. However, it still
yields non-associativity, since we are replacing
matrix products, such as $AB$, by $A\cdot B$
times the unit matrix. But in any case, it is evident
that with the $\mhrt$ product the set
${\cal A}$ becomes a closed algebra, which we denote by \cas
\cite{f3}.

\subsection*{Complex numbers from real}

Before continuing, let us consider
the simplest example of the above matrices,
namely the case $N=1$. In this circumstance, the
matrices $A$ and $B$ become simple {\em
commuting} numbers, $a$ and $b$.
If we specialize further, and choose
$\beta=\al$ and $b=-a$ to be real, we end up with
2-parameter matrices. Their products are
\be
X~\mhrt~ X'\equiv
\ttc{\al} {a} {-a} {\al} \mhrt
\ttc{\al' }{a' }{-a' }{\al'} :=
\ttl{\al\al'-\laz aa'}
{~~\al a'+ \al' a}
{\al' a+ \al a'}
{~~\al \al' -~\laz aa'},
\el{mrc}
and this is nothing but the
multiplication rule of two complex numbers
$z$ and $z'$, provided that we set $\laz=1$
and identify $\al$ and $a$ with the real and
imaginary parts of $z~$.
Thus, a {\em subalgebra} of $\cas$ for $N=\lambda_0=1$
becomes isomorphic to the complex numbers $\CC$:
\be
\cas \ni X =
\ttc{\al} {a} {-a} {\al} \Lra z\equiv \al +i a \in
\CC
\el{mrrr}.

\subsection*{Simple and Hermitian conjugates}

The attractive feature of the generalization (53)
is that most results and definitions needed for
octonions apply almost automatically to $\cas$.
For example, for every element $X \in \cas$
we can define a
conjugate element $\wb{X}$, as follows:
\be
\wb{X} =\overline {\ttl{\al} {A} {B} {\beta}}:=
\ttc{\beta} {-A} {-B} {\al}~.
\ee
By substituting $A'\ra -A~$,
$B'\ra -B$, $\al'\ra \beta$
and $\beta'\ra \al$
in (53), we get immediately
\be
X \mhrt \wb{X}=
{\ttl{\al} {A} {B} {\beta}} \mhrt
\ttc{\beta} {-A} {-B} {\al}=
(\al\beta - \laz~ A \cdot B)
\ttc{1} {0} {0} {1} \equiv n(X) I_{2N\x 2N}~,
\el{norm}
where $n(X) \in \CC~$.
In the meantime, we should again look upon $n(X)$
simply as a scalar function, defined by the
map $\cas \ra \CC$~ in (58). Later we shall
study the conditions on $\cas$ under which $n(X)$
becomes a norm.

\section*{V. Subalgebras of \cas }

The algebra \cas has several interesting
subalgebras~:

\begin{itemize}

\item
An obvious subalgebra is the one
obtained by choosing both matrices $A$ and
$B$ to be traceless:
\be
{\cas_0 }:= \left\{ \ttc{\al}{A}{B}{\beta}
~\vline
\tr A=\tr B =0 \right\} ~,
\ee
\item
This subalgebra has in turn another
subalgebra $\cas_A \subset \cas_0~$,
in which $A$ and $B$ become antisymmetric matrices.
\item
A third subalgebra, which we denote by \casp, is
obtained by choosing $\beta
=\al$ and $B=-A$~:
\be
{\casp }:= \left\{ \ttc{\al}{A}{-A}{\al}
\right\} ~.
\ee
It is easily verified that products of such matrices
stay in the same class:
\br
X~\mhrt~X'&=&\ttc{\al}{A}{-A}{\al} \mhrt
\ttc{\al'}{A'}{-A'}{\al'} \nn \\
&=&
\ttl{\al\al'-\laz~A\cdot A'}
{\al A'+\al'A-\la [A,A']}
{-\al A'-\al'A+\la [A,A']}
{\al\al'- \laz~A\cdot A'} \in \casp~.
\er
Moreover, \casp has the interesting property:

\prop{1} {\em The subalgebra \casp is flexible
for all matrices $A$~}.

To put this result into perspective, we
note that all abelian or anti-commutative algebras are
flexible; thus if  $yx=\pm xy$, then
$x(yx)=\pm (yx)x=(xy)x$, so that
$(x,y,x)=0~.$
Therefore, it is of interest to show that
\casp, which is neither
abelian nor anti-commutative, is also
flexible. \\

\pf We shall prove the above assertion by
explicit multiplication. However, to simplify
the calculations we first note that the
multiples of unity added to each element do not affect the
associators:
\be
(X+\al 1,Y+\beta 1,Z+\gamma 1)= (X,Y,Z)~,
\el{ind}
where $1$ is the identity matrix. This follows
immediately from the linearity of associators:
\be
(X+\al 1,Y,Z)=
(X,Y,Z)+ \al (1,Y,Z)=
(X,Y,Z)~.
\ee
The property \rf{ind} is helpful for calculating associators of
the subalgebra \casp, since we can set the $\al$'s
equal to zero, when calculating the associators.

We now calculate
explicitly the associator $(X_1,X_2,X_3)~$
for general matrices from \casp, but using
only those with $\al_i=0$, {\em i.e.}
\be
X_i =\ttm{0}{A_i}{-A_i}{0}\in \casp~, \for
i=1,2,3~.
\ee
We get
\be
(X_1,X_2,X_3) \equiv
(X_1X_2)X_3 - X_1(X_2X_3)\!=\!
\ttc{p}{P}{-P}{p}-\ttc{q}{Q}{-Q}{q}\!=\!
\ttc{p-q}{P-Q}{Q-P}{p-q},
\ee
where
\br
p &=& \la \laz~([A_1,A_2]\cdot A_3)~, \\
P &=& -\laz (A_1\cdot A_2)~ A_3
+ \la^2 [[A_1,A_2],A_3]~.
\er
and
\br
q&=& \la\laz~(A_1\cdot [A_2, A_3])~, \\
Q &=& -\laz (A_2\cdot A_3)~ A_1
+ \la^2 [A_1,[A_2,A_3]]~.
\er
Therefore, the elements of the associator
$(X_1,X_2,X_3)$ are
\br
p-q &=& \la \laz ~\left([A_1,A_2]\cdot A_3- A_1\cdot
[A_2, A_3] \right)=0~, \\
P-Q &=&
-\laz \left( (A_1\cdot A_2)~ A_3
- (A_2\cdot A_3)~ A_1\right)
+ \la^2 \left( [~[A_1,A_2],A_3]
- [A_1,[A_2,A_3]~]\right) \nn \\
&=&
-\laz
\left( (A_1\cdot A_2)~ A_3
- (A_2\cdot A_3)~ A_1\right)
+ \la^2 [A_2,[A_3,A_1]~]~.
\erl{con7}
In other words, {\em the associator
$(X,Y,X)$ vanishes identically, for any
$\la,\laz, A_1=A_3 {\rm~and~} A_2 $},
\be
(X,Y,X)=0~, \for X,Y \in \casp~.
\el{flex2}
\qed

\item As a fourth subalgebra,
let $\gb$ be a given Lie algebra of dimension $n$,
and let $V_\gb$ be the algebra spanned by the
representation matrices of \gb. Then, we can define
a subalgebra of \cas via
\be
\gbd:= \left\{ \ttc{\al}{A}{B}{\beta}
~\vline
A,B \in V_\gb \right\} ~.
\ee
Clearly the off-diagonal elements,
such as $\al A'+\beta' A+\la [B,B']$,
of the products $X\mhrt X'$ belong to $V_\gb$~.
Hence, $\gbd$ are subalgebras of $\cA_0$.
Moreover, half of $\gbd$~, obtained by the
intersection of $\gbd$ with \casp, will be a
subalgebra of $\gbd$:
\be
\gb^D_0:= \left\{ \ttc{\al}{A}{-A}{\al}
~\vline
A \in V_\gb \right\} \subset \gbd \subset \casp ~.
\ee
The commutators of the elements of $\gbdz$ constitute
a Lie algebra, which is isomorphic to the original
algebra \gb~.

\end{itemize}

\section*{VI. Grading of \cas }

\prop{2} {\em The algebra $\cA^{\mhrt}$ can be graded,
as follows}:
\be
\cas = \casp \oplus \casn
= \casp \oplus K\casp= \casp \oplus K \mhrt \casp~,
\ee
{\em where the `grading matrix' is}
\be
K\equiv \ttc{1}{0}{0}{-1}~.
\ee
Observe that $K\mhrt X = KX$ for any $X \in A^\hrt$. Also of
course
\be
\A^\hrt_\eta ~\mhrt~ \A^\hrt_{\eta'} \subseteq
\A^\hrt_{\eta\eta'}~.
\el{incl}

\pf Every matrix $X\in \cas$ can be
decomposed, as follows:
\br
X &\equiv& \ttc{\al}{A}{B}{\beta}=
\ttc{\al_+}{A_+}{-A_+}{\al_+} +
\ttc{\al_-}{A_-}{A_-}{-\al_-}
\equiv X_+ ~+~ \widehat X_- \lb{de} \\
&\equiv& X_+ ~+~ KX_-
\erl{dec}
where
\be
\al_\pm \equiv  \hf(\al \pm \beta)~,\quad
A_\mp \equiv\hf (A \pm B)~, \quad
K\equiv \ttc{1}{0}{0}{-1}~,
\ee
The first set of matrices (with
$\beta=\al$ and $B=-A$) constitutes the subalgebra
\casp, which we defined earlier in
(59). The second set of matrices
(with $\beta=-\al$ and $B=A$) will be
called \casn.
Since $\casp$ is a subalgebra of $\cas$~,
clearly $\casp \casp =\casp$~.
The rest of the inclusion relations \rf{incl},
namely
\be
\hat X\mhrt \hat X'\in \casp~, \quad
X\mhrt \hat X'\in \casn~, \quad
\hat X\mhrt X'\in \casn~.
\ee
follow immediately from the equalities (85)~-~(87)
which we shall prove below. \qed
\vspace{.2in}

\prop{3} {\em The following equalities hold for
any $X,X'\in \casp$}:
\br
KXK &=& \wbar{X}~, \\
(KX) \mhrt (KX') &=& X' \mhrt \wbar{X}~, \qquad \\
X \mhrt (KX') &=& K(\wbar{X}\mhrt X')~, \qquad \\
(KX) \mhrt X' &=& K(X' \mhrt X)~. \qquad \lb{kr}
\er

\pf The proof follows simply by explicit
matrix multiplication, using (60):
\br
KX ~\mhrt~ KX'\!\! &\equiv&\!\!
\ttc{\al}{A}{A}{-\al}\! \mhrt\!
\ttc{\al'}{A'}{A'}{-\al'}
=
\ttl{\al\al'+ \laz~A\cdot A'}
{\al A'-\al'A-\la [A,A']}
{-\al A'+\al'A+\la [A,A']}
{\al\al'+ \laz~A\cdot A'} \nn \\
&=&
\ttc{\al'}{A'}{-A'}{\al'} \mhrt
\ttc{\al}{-A}{A}{\al} \equiv
X' ~\mhrt~ \bar X \in \casp~. \\
X \mhrt(KX')\!\! &=&\!\!
\ttc{\al}{A}{-A}{\al}\! \mhrt\!
\ttc{\al'}{A'}{A'}{-\al'}
=
\ttl{\al\al'+\laz~A\cdot A'}
{\al A'-\al'A+\la [A,A']}
{\al A'-\al'A+\la [A,A']}
{-\al\al'- \laz~A\cdot A'} \nn \\
&=&\!\!
\ttc{1}{0}{0}{-1}\left\{ 
\ttc{\al}{-A}{A}{\al}\! \mhrt\!
\ttc{\al'}{A'}{-A'}{\al'}\right\}= 
K(\bar X\! ~\mhrt~\!X') \equiv
\widehat{\bar X\!~\mhrt\!~X'} \in \casn~, \\
KX \mhrt X' &=& 
\ttc{\al}{A}{A}{-\al} \mhrt 
\ttc{\al'}{A'}{-A'}{\al'} = 
\ttl{\al\al'-\laz~A\cdot A'}
{\al A'-\al'A+\la [A,A']}
{\al A'-\al'A+\la [A,A']}
{-\al\al'- \laz~A\cdot A'} \nn \\ 
&=& 
\ttc{1}{0}{0}{-1}\left\{ 
\ttc{\al'}{A'}{-A'}{\al'} \mhrt
\ttc{\al}{A}{-A}{\al}\right\}= 
K(X' ~\mhrt~X) \in \casn~.
\er

\subsection*{Matrix representation of the 
Cayley-Dickson process}

If we multiply the grading matrix $K$ by a real or
complex scalar $v$, and let $\mu \equiv v^2$~, we get 
\be
v_{op} := vK~, \qquad 
v_{op}v_{op} := v^2 1=\mu 1~. 
\ee
Therefore, using the relations (81)~-~(84), we get the 
multiplication rule  
\be
(X_1 +v_{op} X_2) \mhrt (X_3 
+v_{op} X_4)=
(X_1\mhrt X_3+\mu X_4 \mhrt \wbar {X_2})+ v_{op}
({\wbar
X_1} \mhrt X_4 + X_3 \mhrt X_2)~, \quad \forall X_i\in
\casp~.
\ee
This is exactly the multiplication rule given by Cayley and Dickson
where one starts with an abstract algebra $\cB$ and defines an
abstract operator $v_{op}$, and essentially {\em postulates} the
following multiplication rule \cite{schaf}:
\be
(b_1 +v_{op}b_2)(b_3 +v_{op}b_4)=
(b_1b_3+\mu b_4 \bar b_2)+ v_{op}(\bar b_1b_4 +
b_3b_2)~, \quad b_i \in \cB~.
\ee
where $\bar b_i\in \cB$ is the conjugate of $b_i$,
and $ v_{op}\notin \cB$~, such that $v_{op}^2=\mu \cdot 1$~.

Observe that the $\mhrt$ multiplication rule provides
an explicit matrix representation of the
Cayley-Dickson process \cite{schaf},
provided that the original
algebra $\cB$ can be represented by \casp.

\subsection*{Composition algebras from
$2\x 2$ matrices}

One may wonder what happens if we allow the rudim
entary
$2\x 2$ matrices to contain arbitrary complex elements.
Since
\be
X ~\mhrt~ X'\equiv {\ttc{\al} {a} {b} {\beta}} ~\mhrt~
{\ttc{\al' }{a' }{b' }{\beta' }} :=
\ttl{\al\al'+ \laz~ ab'} {~~\al a'+ \beta' a}{\al' b+ \beta b'}
{~~\beta\beta'+ \laz~ ba'},
\el{mro}
when $X'=\wbar{X}$ this product yields a `norm',
\be
X*\wbar{X}=n(X)\ttc{1}{0}{0}{1}, \where
n(x)= \al\beta -\laz ab~.
\el{normo}
It is easy to check, by explicit multiplication,
that the following identity holds for any
$\la_0\in \CC$:
\be
n(x)n(x')=
(\al\beta -\laz ab)(\al'\beta' -\laz a'b')=
(\al\al'+ \laz~ ab')(\beta\beta'+ \laz~ ba')
-\laz (\al a'+ \beta' a) (\al' b+ \beta b')
\!=\!n(xx')
\ee
which informs us that the standard norm
\rf{normo}
for $N=1$ obeys the composition law.

Clearly, the norm (92) is degenerate for any
$\la_0\in \CC$, if we allow $X$ to be any $2\x 2$
matrix. (For example, simply choosing $\beta=a=b=1$
and $\al=\laz$ will yield an $x\ne 0$ with
$n(x)=\al\beta - \laz ab=0~$.)
The question then arises, when is the norm $n(x)$
in \rf{normo} nondegenerate?
We can certainly guarantee that $n(x)$ is
nondegenerate,
if we restrict $X$ to have the following special
form:
\be
X=\ttc {z}{w}{-\bar w}{\bar z}, \aand \Re \la_0>0~.
\el{condn}
whence
\be
n(x)= |z|^2+\laz |w|^2 \ne 0~, \for \Re \la_0\ne 0~.
\ee
[Of course, there exist a few equivalent variations of
the conditions \rf{condn}.
For instance, we can replace $-\bar w$ by $\bar w$,
but demand that $\Re \laz <0~$.]

Anyhow, this means that we are dealing with a division algebra,
which must therefore be one of the 4 possibilities.
Because $X\in M_{2\x 2}$~, we may expand it in
terms of Pauli matrices, getting
\br
X &=&
\ttc {z}{w}{-\bar w}{\bar z}=
z_1 \ttc {1}{0}{0}{1}+
i w_2 \ttc {0}{1}{1}{0}+
i w_1 \ttc {0}{-i}{i}{0}+
i z_2 \ttc {1}{0}{0}{-1} \nn \\
&\equiv& x_0 \s_0-i {\bf x}\cdot \sv~\leftrightarrow ~x_0e_0+
 \sum_{k=1}^3 x_k e_k~,
\where w_i,z_i\in \RR~
\erl{qr}
Since $e_0 \ra \s_0$ and $e_j \ra -i\s_j ~j=1,2,3~$
are known representations of the quaternions,
we conclude that this is the quaternion algebra
over the {\em real} field $\RR$~, as expected.
Indeed the matrix \rf{qr} is the usual representation of
${\cal Q}$ in terms of standard matrices. Later on we shall
describe another representation by nonstandard matrices.

\section*{VII. Conditions on deformation parameters}

Previously we showed that \casp is flexible.
We now ask under what conditions \casp can become
alternative.

For this, we must have $(X_1,X_1,X_3)=0~.$
By setting  $X_2=X_1$ and noting Eq.\rf{con7}, we get the condition
\be
[A_1,[A_3,A_1]~] = \frac{\la^2}{\laz}
=\left((A_1\cdot A_1)~ A_3 - (A_1\cdot A_3)~ A_1\right)~.
\ee
This condition can be satisfied if
$\la^2=\laz/4$ and the $A_i=\ab_i\cdot \sv$~:
Indeed, for such $A_i$, we get
\br
[A_2,[A_3,A_1]~] &=& [\ab_2\cdot\sv,
2i[\ab_3\x \ab_1]\cdot \sv ]
= -4 [ \ab_2\x [\ab_3\x \ab_1]]\cdot \sv
\nn \\ &=& 4
((\ab_2\cdot\ab_3) \ab_1-
(\ab_2\cdot\ab_1) \ab_3) \cdot \sv
=4\left((A_1\cdot A_2)~ A_3
- (A_2\cdot A_3)~ A_1\right)
\er
where we used $(A\cdot B)=\frac 1 2 \tr (AB)$~.
Hence, we can have $\la=\pm \hf \sqrt{\laz}$~.
For the special choice $\laz=-1$, we get
$\la=\pm \frac i 2 $~.

This sign ambiguity is the
origin of the non-uniqueness of the $\mhrt$ product \cite{f4}.

\section*{VIII. Summary}

Our algebras provide concrete matrix
representation of a big class of non-associative
algebras. They may suggest new constructions in the future.
One such possibility, which leads to
notions of triality, is described in the Appendix A, but anyhow
the formulation (18) and its deformations (53)
permit generalizations that have obvious affinity
to higher symmetry groups, rather than the simple case of SU(2).
We also believe that our treatment of hermiticity and norm
for the complex case is reasonable; we
illustrate their utility with reference to the Dirac
equation in Appendix B.

\section*{Acknowledgements}
We are extremely grateful to Dr G Joshi for supplying us with
a comprehensive list of references in this vast topic and for
pointing us in the right direction, and to Dr B Gardner
for useful comments.


\section*{Appendix A. The $\di$ product and triality}

In this appendix we try another type of product, which we denote by
$\di$, where the
commutators $[B,B']$ in the $\hrt$ product are now replaced by the
standard matrix products $BB'$.

Let us first consider the simplest case, $N=1$,
where the matrices $A$ and $B$ become scalars, so that we shall
first deal with $2\times 2$ matrices:
\be
X \equiv \ttc{\al} {a} {b} {\beta}.
\ee
We define the new matrix product, as follows
\be
X ~\di~ X'\equiv {\ttc{\al} {a} {b} {\beta}} \di
\ttc{\al' }{a' }{b' }{\beta' } :=
\ttl{\al\al'+ \laz~ ab'}
{~~\al a'+ \beta' a +\la bb'}
{\al' b+ \beta b'+ \eta \la aa'}
{~~\beta\beta'+ \laz~ ba'},
\el{mrss}
where $\eta,\la, \la_0$ are arbitrary complex numbers. We now ask the
question, whether for such a product, we can define for every $X$ a
conjugate $\wbar{X}$, such that
 $X\di \wbar{X} = n(X) \cdot 1$~, where $n(X)$ is some quadratic form
of $X$, i.e. $n(sX)=s^2n(X)$~.

Let us try the following ansatz:
\be
\wbar{X} \equiv
\ttc{\beta+\gamma}{-a}{-b}{\al+\delta}~.
\ee
We want to determine $\gamma$ and $\delta$,
and derive conditions on
$\al,\beta,a,b$,
by demanding that $X\di \wbar{X} \propto 1$~:
\br
X \di \wbar{X} &=&
\ttl{\al}{a }{b}{\beta} \di
\ttc{\beta+\gamma}{-a}{-b}{\al+\delta}
= \ttl{\al(\beta+\gamma)- \laz ab }
{\delta a-\la b^2 }{\gamma b-\eta \la a^2}
{\beta(\al+\delta)- \laz ab } \nn \\
&=& n(X)\ttl{1}{0 }{0 }{1 }~,
\where n(X)=\al(\beta+\gamma)- \laz ab~.
\erl{con3}
This condition is obeyed, if
\be
\delta a-\la b^2=0~, \qquad  \gamma b-\eta \la a^2=0~,
\aand \alpha\gamma = \beta\delta,
\ee
We cannot satisfy these conditions for
general $a$ and $b$. (For example, if $a=0$ and
$b\ne 0$, then $\delta$ would be infinite.)
Thus, for $\eta\ne 0$ we must assume that either both
$a$ and $b$ are zero or both
are unequal to zero.
But for $\eta=0$ we must demand $b=0$. With these restrictions, we
get
\be
\delta(X)= \la \frac{b^2}{a}~,
\qquad
\gamma(X)= \eta \la \frac{a^2}{b}~,
\aand
\frac \beta \al=\frac \gamma \delta
=\eta \frac{a^3}{b^3}~.
\el{con2}
An additional condition can be obtained by demanding
that the adjoint operation is an
involution, so that $\wbar{\wbar{X}}=X$,
whence
\br
\wbar{\wbar{X}} &=&
\ttl{\al+\delta(X)+\gamma(\wbar{X})}
{a}{b}{\beta+\gamma(X)+\delta(\wbar{X})} \nn \\
&=&
\ttl{\al+\la(\frac{b^2}{a}-\eta \frac{a^2}{b})}
{a}{b}{\beta+\la(\eta \frac{a^2}{b}+\frac{b^2}{-a})}=
\ttl{\al}{a}{b}
{\beta}=X~.
\er
Thus, we get the new condition
\be
b^3=\eta ~a^3~, \orr b=\xi a~, \where \xi=\eta^{1/3}
~,
\el{ac}
Note that for each $\eta$ we have three cubic roots $\xi$~.
Substituting
\rf{ac} into the third equality in \rf{con2}, we get
\be
\alpha = \beta ~, \aand
\gamma=\delta~.
\ee
Hence, $a$ can be any complex number as long as
$b=\xi a$.
Finally, by noting all the above conditions, we
get for every $\eta \in \CC$ three
sets of $2\x 2$ matrices, which
are closed algebras under the product
$\di$~:
\be
X(\xi) = \left\{ \ttl{\al} {a} {\xi a} {\al} \right\}~,\qquad \xi
=\eta^{1/3} \in
\CC~,
\ee
For these matrices, the adjoint and the corresponding quadratic form are
\be
\wbar{X} \equiv
\ttc{\al + \xi^2 \la a}{-a}{-\xi a}{\al+\xi^2 \la a }~.
\ee
and
\br
n(X) &=& \al(\beta+\gamma)- \laz ab=
\al^2 + \eta \la~ \al \frac {a^2} b - \laz ab \nn \\
&=&
\al^2 + \xi^2 \la~ \al a -\xi \laz a^2 ~.
\er
Note that $n(X)$ is quadratic in $X$, {\em i.e.}
$N(sX)=s^2~N(X),\, s\in \CC~$.

For the special case $b=-a$, or  $\xi=-1$~, we obtain a known
quadratic form \cite{zsss}
\be
n(X)=\al^2 +\la~ \al a +\laz a^2~.
\ee
Proceeding to larger matrices,  let
\be
X \di X'\equiv {\ttc{\al} {A} {B} {\beta}} \di
{\ttc{\al' }{A' }{B' }{\beta' }}\! :=\!
\ttl{\al\al'+ \laz~ A\cdot B'}
{~~\al A'+ \beta' A + \la BB'}
{\al' B+ \beta B'+\eta\la AA'}
{~~\beta\beta'+ \laz~ B\cdot A'},\quad \eta \in \CC~.
\ee
One can readily check that such matrices yield closed algebras
with respect to the above $\di$ product. By restricting $B$ to be $\xi A$
we get the following subalgebra:
\be
X(\xi) = \left\{ \ttl{\al} {A} {\xi A} {\al}
\right\}~, \qquad \xi=\eta^{1/3}\in \CC~.
\ee
We can check, using (112), that products of two such matrices yield a
matrix of the same type:
\be
X \di X'\equiv{\ttc{\al}{A}{\xi A}{\al}}\di{\ttc{\al'}{A'}{\xi A'}{\al'}}
\! =\!\ttl{\al\al'+ \xi\laz~ A\cdot A'}{~~\al A'+ \al' A + \xi^2 \la AA'}
{\xi (\al' A+ \al A')+\eta\la AA'}{~~\al\al'+ \xi \laz~ A\cdot A'}.
\ee
However, if we replace the {\em scalar} $a$ in eqs. (109) and (110)
by a {\em matrix} $A$, we do not get an adjoint nor a bilinear form, since
the appropriate items do not stay scalar, as they should.

Finally we note that if we replace the simple products $AA'$ in (114) by
anticommutators $\{A,A'\}/2$ and if we make the scalar products $A\cdot
A'$ symmetric, i.e.
$$
{\ttc{\al}{A}{\xi A}{\al}}\di_S~{\ttc{\al'}{A'}{\xi
A'}{\al'}}
\! :=\!\ttl{\al\al'+ \xi\laz~ A\cdot A'}{~~\al A'+ \al' A + \xi^2
\la \{A,A'\}/2}
{\xi (\al' A+ \al A')+\eta\la
 \{A,A'\}/2}{~~\al\al'+ \xi \laz~ A\cdot A'}~,
$$
then the product becomes abelian. Therefore the new algebra will
automatically become {\em flexible}.

One can similarly symmetrize the more general product (112) by defining,
$ X \di_S~ X' := (X \di X' +X' \di X)/2$, and also get a flexible
algebra.

\section*{Appendix B. Application to the Dirac equation}

In momentum space, the free Dirac equation reads
\be
P\Psi \equiv (p_0-{\bf p}\cdot\alb - m\beta)\Psi =
\left( \begin{array}{cc}
       p_0-m & -{\bf p}\cdot{\sv}\\
       -{\bf p}\cdot{\sv} & p_0+m
       \end{array} \right) \Psi   = 0~.
\el{deq}
By noting that the standard matrix product and the $\mhrt$
products between the Dirac operator
$P$ and its conjugate $\bar P$ are equal, we immediately
obtain
\be
P  \bar P= P\mhrt \bar P = n(P)~I = \left((p_0-m)(p_0+m)-{\bf p}^2\right)~
I=0~,
\ee
where we used Eq. \rf{ml} to calculate the norm $n(P)$.
Therefore each of the 4 columns of $\bar P$ will be a solution of the
Dirac equation \rf{deq}. Hence, we write
\be
   \Psi =\bar{P} = \left(
  \begin{array}{cc} p_0+m & {\bf p}\cdot{\sv}\\
     {\bf p}\cdot{\sv} & p_0-m \end{array} \right).
\ee
The first and second columns of $\Psi$ are proportional to the positive
energy solutions $u^1(p)$ and $u^2(p)$, while the third and fourth
columns yield the negative energy solutions $v^1(p)$ and $v^2(p)$, if we
replace $m$ by $-m$, because
\be
  (p_0-{\bf p}\cdot\alb - m\beta)u^i(p) =0, \qquad
  (p_0-{\bf p}\cdot\alb + m\beta)v^i(p) =0,\quad
  (i=1,2).
\ee
Therefore the {\em physical and normalizable} solutions can be expressed
as
\be
   \Psi_{ph} \equiv \left( u^1|u^2|v^1|v^2 \right) =
   \frac{1}{\sqrt{2m(p_0+m)}}\left( \begin{array}{cc}
   p_0+m & {\bf p}\cdot{\sv}\\
   {\bf p}\cdot{\sv} & p_0+m \end{array}\right)~.
\ee
The orthogonality and normaliztion relations among these solutions
\br
\bar{u}^i(p)v^j(p)\equiv u^{i\dag}(p)\beta v^j(p)=0~~, \quad
&{\rm and}&
\quad \bar{v}^i(p)u^j(p)\equiv v^{i\dag}(p)\beta
u^j(p)=0~, \quad (i,j=1,2), \nn \\
\bar{u}^i(p)u^j(p)\equiv u^{i\dag}(p)\beta
u^j(p)=\delta_{ij}, \quad
&{\rm and}&
\quad \bar{v}^i(p)v^j(p)\equiv v^{i\dag}(p)\beta
v^j(p)=-\delta_{ij} \nn
\er
can also be elegantly summarized and proved using the matrix
representations of
octonions, as follows
\be
\Psi_{ph}^\dag \beta \Psi_{ph}=
\Psi_{ph} \beta \Psi_{ph}=
\Psi_{ph} \bar {\Psi}_{ph} \beta =
\left( \Psi_{ph} \mhrt \bar {\Psi}_{ph}\right) \beta =
n(\Psi_{ph}) \beta = \beta~.
\ee
Finally, it is interesting to note that the Dirac operator $P$ is equal
to our matrix representation of the following octonion:
\be
P \equiv (p_0-{\bf p}\cdot\alb - m\beta) =
\left( \begin{array}{cc}
       p_0-m & -{\bf p}\cdot{\sv}\\
       -{\bf p}\cdot{\sv} & p_0+m
       \end{array} \right)\Leftrightarrow
p \equiv (p_0 + i{\bf p}\cdot{\bf \hat{e}} + ime_4)~,
\ee
where we have used the correspondence
\be
\alpha_k = \left( \begin{array}{cc} 0 & \sigma_k\\
                  \sigma_k & 0 \end{array} \right)
         = -i\hat{\Omega}_k \Leftrightarrow -i\hat{e}_k, \qquad
\beta = \left( \begin{array}{cc} 1 & 0 \\
               0 & -1 \end{array} \right)
         = -i\Omega_4 \Leftrightarrow -ie_4.
\ee
Notice that we cannot write the Dirac equation in terms of quaternions alone,
since we require five different basis elements: $(e_0, e_4,\hat{e}_k;\,
k=1,2,3)$ or $(e_0, e_4, e_k;\, k=1,2,3)$.
Also note that
the octonion $p$ is nicely hermitian ($p=p^\dag$)~.


\begin{thebibliography}{99}
\bibitem{f2} Note, that although
$X_i\cdot X_j=-2\delta_{ij}~~(i,j=1,2,3)$
for $X_i\in so(3)$, the diagonal elements of
$X_i\cdot X_j$ can have different signs, as in
$so(2,1)$, where $(X_i\cdot X_j)= \diag(-2,-2,2)$~.
\bibitem{f3} Actually we can generalize
\cas further, by replacing the product of the complex numbers
$\al\beta$ with a bilinear product
$\al \beta \longleftrightarrow
\al \cdot \beta = \al_+^2 + \la_1 \al_-^2~,$
where $\al_\pm=(\al \pm \beta)/\sqrt{2}.$ We have refrained
from doing so in the text.
\bibitem{f4} Thus if we reverse the sign in front of the $\sigma$
in (17) we only need to change the sign of the commutators
in (18) to fix the multiplication rule.
\end{thebibliography}
\end{document}